\documentclass[]{style/ceurart}

\usepackage[raggedrightboxes]{ragged2e}

\usepackage{amsmath}
\usepackage{makecell}
\usepackage{bbm}

\begin{document}

\copyrightyear{2024}
\copyrightclause{Copyright for this paper by its authors.
  Use permitted under Creative Commons License Attribution 4.0
  International (CC BY 4.0).}

\conference{CLEF 2024: Conference and Labs of the Evaluation Forum, September 9-12, 2024, Grenoble, France}

\title{Transfer Learning with Pseudo Multi-Label Birdcall Classification for DS@GT BirdCLEF 2024}

\author[1]{Anthony Miyaguchi}[
orcid=0000-0002-9165-8718,
email=acmiyaguchi@gatech.edu,
url=https://linkedin.com/in/acmiyaguchi,
]
\cormark[1]

\author[1]{Adrian Cheung}[
orcid=0009-0006-8650-4550,
email=acheung@gatech.edu,
url=https://linkedin.com/in/acheunggt,
]
\cormark[1]

\author[1]{Murilo Gustineli}[
orcid=0009-0003-9818-496X,
email=murilogustineli@gatech.edu,
url=https://linkedin.com/in/murilo-gustineli,
]
\cormark[1]

\author[1]{Ashley Kim}[
email=akim614@gatech.edu,
url=https://www.linkedin.com/in/-ashleykim/,
]

\address[1]{Georgia Institute of Technology, North Ave NW, Atlanta, GA 30332}
\cortext[1]{Corresponding author.}

\begin{abstract}
    We present working notes for the DS@GT team on transfer learning with pseudo multi-label birdcall classification for the BirdCLEF 2024 competition, focused on identifying Indian bird species in recorded soundscapes. 
    Our approach utilizes production-grade models such as the Google Bird Vocalization Classifier, BirdNET, and EnCodec to address representation and labeling challenges in the competition.
    We explore the distributional shift between this year's edition of unlabeled soundscapes representative of the hidden test set and propose a pseudo multi-label classification strategy to leverage the unlabeled data.
    Our highest post-competition public leaderboard score is 0.63 using BirdNET embeddings with Bird Vocalization pseudo-labels.
    \break
    Our code is available at \href{https://github.com/dsgt-kaggle-clef/birdclef-2024}{github.com/dsgt-kaggle-clef/birdclef-2024}.
\end{abstract}

\begin{keywords}
    Transfer Learning \sep
    Dataset Annotation \sep
    Embeddings \sep
    Association Rule Mining \sep
    Google Bird Vocalization Classifier \sep
    BirdNET \sep
    EnCodec \sep
    CEUR-WS
\end{keywords}

\maketitle

\section{Introduction}

The BirdCLEF 2024 competition \cite{birdclef2024} involves identifying bird species in 4-minute-long soundscapes recorded in the Western Ghats of India as part of the LifeCLEF lab \cite{lifeclef2024}.
Passive monitoring of ecological areas allows humans to determine how to allocate our attention to preserve biodiversity for posterity.
The primary method of passive monitoring for birds is through bioacoustics.
Given a recording of a soundscape located on an autonomous recording device in the field, we would like to determine when and where specific birds are vocalizing.

The objective of the BirdCLEF competition is to predict the presence of each of the 182 target species for every 5-second segment in test soundscapes.
One of the main challenges this year is the 4400 minutes of test soundscapes that must be predicted in 120 minutes of CPU time instead of 2000 minutes in BirdCLEF 2023 \cite{kahl2022overview}.
We focus on transfer learning using Google’s Bird Vocalization Classification Model, made publicly available in the BirdCLEF 2023 competition.
We explore embeddings from BirdNET, a model trained on bird vocalizations, and EnCodec, a neural audio codec trained on diverse audio data.
\section{Birdcall Classification Overview}

Birdcall classification is a challenging task due to the variability in bird vocalizations, the presence of background noise, and the large number of species to classify.
In addition, the measured data comes in the form of audio recordings, which are high-dimensional and require specialized processing techniques.
Many successful approaches to birdcall classification utilize convolutional neural networks (CNNs) to extract features from audio spectrograms.
Audio spectrograms are a time-frequency representation of the audio signal, which are extracted using the short-time Fourier transform (STFT) \cite{durak2003short}, often with additional preprocessing steps such as mel-frequency scaling.
The spectrograms are represented as 2D images, which utilize techniques in the rich literature surrounding image classification.

BirdNET is a popular birdcall classification model that utilizes the spectrogram-CNN approach.  
It is widely distributed in the field due to its high accuracy and ease of use on mobile devices \cite{kahl2021birdnet}.
The Google Bird Vocalization Classifier is another model using EfficientNet-B1, a similar CNN architecture, and is trained on many soundscapes.
It was released alongside the BirdCLEF 2023 competition and has more than 10,000 species in its output space \cite{denton2022birdvocalization}.

\section{Domain Knowledge Transfer via Embedding Spaces}

Neural networks are universal function approximators that map some input space to an arbitrary output space \cite{lecun2015deep}.
A neural network's intermediate layers can be considered a manifold that typically projects high dimensional data to lower dimensional spaces.
Transfer learning is a technique that leverages the learned representations of a model trained on one task to improve the performance of a model on a different task.
In the context of birdcall classifications, raw audio data is projected onto a manifold optimized to discriminate between bird species.
The projections are called embeddings and can be used as features in downstream tasks to transfer knowledge to a new but similar domain.
Few-shot learning on global bird embeddings tends to be effective on new domains as per \citet{ghani2023global}.

\begin{figure}[h!]
    \centering
    \includegraphics[width=\textwidth]{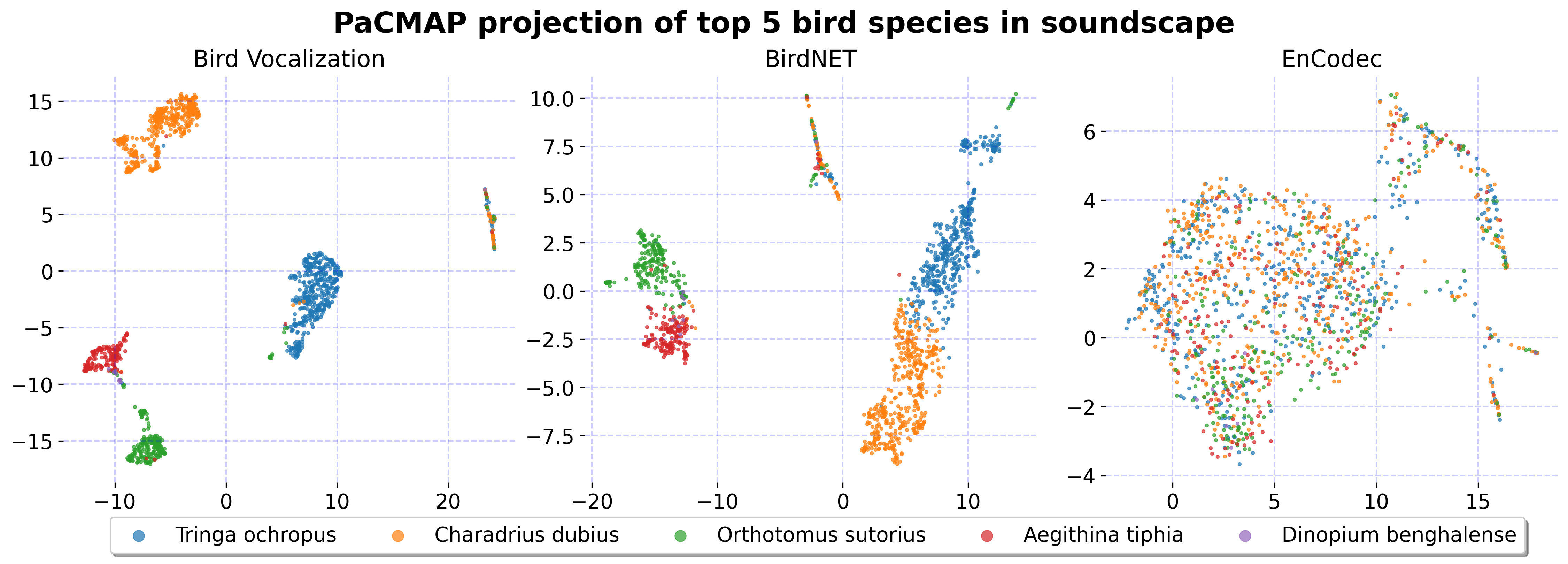}
    \caption{
        PaCMAP projections of the top five species averaged embeddings ranked by soundscape frequency.
        Embeddings can be evaluated qualitatively by clustering behavior.
    }
\label{fig:pacmap-cluster}
\end{figure}

We visually inspect learned embeddings by projecting them into two or three dimensions to reveal meaningful semantic structures that can be qualitatively interpreted by humans.
We use PaCMAP, a dimensionality reduction technique that preserves local and global structure on a manifold via pairwise relationships \cite{wang2021pacmap}, to visualize embeddings from the Bird Vocalization Classifier, BirdNET, and EnCodec.
In Figure \ref{fig:pacmap-cluster}, we project the top five species by frequency in the soundscape dataset.
The first two are domain-specific models trained on bird vocalizations, while the latter is a general-purpose neural audio codec.
EnCodec is particularly interesting because it is not trained on bird vocalizations but rather a diverse set of audio data using a self-supervised learning objective guided by self-attention mechanisms \cite{defossez2022encodec}.
It is also perceptually optimized for audio compression, preserving meaningful structures in the embedding space, such as bird vocalizations over static background noise.

We explore the effectiveness of transfer learning, where bird vocalization predictions are surrogates for true labels.
We hypothesize that transfer learning is an effective technique for the competition because existing models capture underlying structures amenable to optimization by simple linear classifiers.
We quantify how well we can learn domain-specific adaptations between different embedding spaces and how well each of these models is suited to capture the underlying structure of the data.

\section{Exploratory Data Analysis}
\label{sec:eda}

We perform an exploratory analysis on the training and unlabeled soundscape datasets to understand species distribution and their co-occurrence patterns.
We hypothesize a shift in species distribution between the training and unlabeled soundscape datasets due to the differences in the recording environments observed in downstream domain knowledge transfer.
The training dataset contains 182 species obtained from crowd-sourced Xeno-Canto recordings.
Because they are crowd-sourced, the training data is likely biased toward clear and distinct vocalizations that typically occur in isolation.
The soundscape dataset is a collection of 4-minute soundscapes from Western Ghats, India, representative of the hidden test set in the competition \cite{birdclef-2024}.
The soundscape is likely to be more intermittent and contain vocalizations that are less distinct and overlapping due to the lack of human-directed attention to the recording process.

\begin{figure}[h!]
    \centering
    \includegraphics[width=1\linewidth]{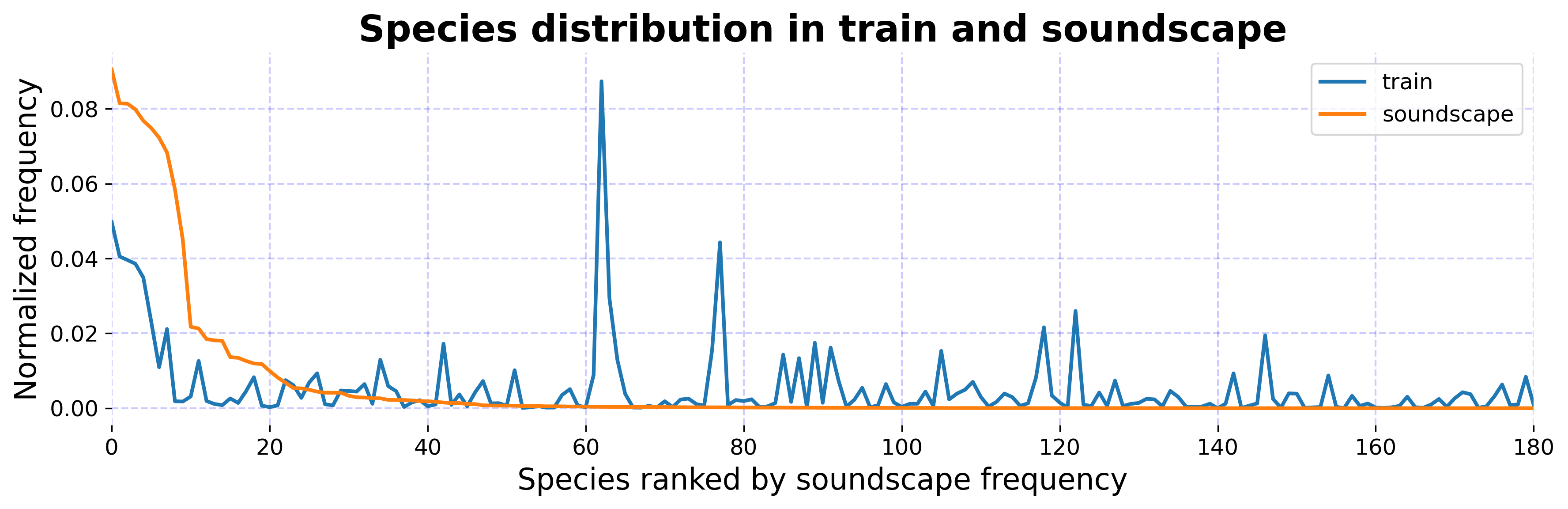}
    \caption{
        Representation of the distribution of species detected in the train and unlabeled soundscapes, sorted by the frequency of species in the soundscape.
    }
    \label{fig:species-distribution}
\end{figure}

We use the Bird Vocalization model to extract the embeddings and logits from the datasets in 5-second intervals.
The training dataset comprises 217k discrete intervals totaling 302 hours, while the soundscape dataset has 407k intervals totaling 566 hours.
We assume that an interval contains a call if the maximum logit value run through a sigmoid function exceeds a threshold of $0.5$.
The training dataset has a higher density of calls, with $62\%$ of intervals containing at least one call, compared to $8.8\%$ in the soundscape dataset.

We compute the relative frequency of species overall discrete intervals and compare their distributions by the ranked frequency in the soundscape dataset in Figure \ref{fig:species-distribution}.
There is a notable discordance between the two distributions, with many species in the training set unrepresented in the unlabeled soundscapes.
We observe that top species in each dataset do not align in Table \ref{table:species_data} using the raw frequency of occurrences.

\begin{table}[h!]
    \centering
    \begin{minipage}{.49\textwidth}
        \caption{Top species for train and soundscape.}
        \begin{tabular}{|c|c|c|c|c|}
            \hline
             & \textbf{Train} & \textbf{Freq} & \textbf{Soundscape} & \textbf{Freq} \\
            \hline
            1 & bncwoo3 & 14170 & grnsan & 6272 \\ \hline
            2 & grnsan & 8069 & comior1 & 5648 \\ \hline
            3 & inbrob1 & 7189 & lirplo & 5638 \\ \hline
            4 & comior1 & 6575 & bkrfla1 & 5532 \\ \hline
            5 & lirplo & 6423 & comtai1 & 5323 \\ \hline
            6 & bkrfla1 & 6260 & btbeat1 & 5189 \\ \hline
            7 & comtai1 & 5664 & putbab1 & 5009 \\ \hline
            8 & houcro1 & 4774 & purher1 & 4735 \\ \hline
            9 & comsan & 4214 & whcbar1 & 4057 \\ \hline
            10 & btbeat1 & 3749 & mawthr1 & 3114 \\ \hline
        \end{tabular}
        \label{table:species_data}
    \end{minipage}
    \hfill
    \begin{minipage}{.45\textwidth}
        \caption{
            Top frequent itemsets in soundscape.
        }
        \begin{tabular}{|c|r|}
            \hline
            \textbf{Items} & \textbf{Freq} \\ \hline
            {[comior1]} & 5320 \\ \hline
            {[lirplo]} & 5308 \\ \hline
            {[lirplo, comior1]} & 5307 \\ \hline
            {[bkrfla1]} & 5283 \\ \hline
            {[bkrfla1, lirplo]} & 5280 \\ \hline
            {[bkrfla1, lirplo, comior1]} & 5280 \\ \hline
            {[bkrfla1, comior1]} & 5280 \\ \hline
            {[grnsan]} & 5197 \\ \hline
            {[grnsan, comior1]} & 5169 \\ \hline
            {[grnsan, lirplo]} & 5168 \\ \hline
        \end{tabular}
        \label{table:itemset_data}
    \end{minipage}
\end{table}

\begin{figure}[h!]
    \centering
    \includegraphics[width=1\linewidth]{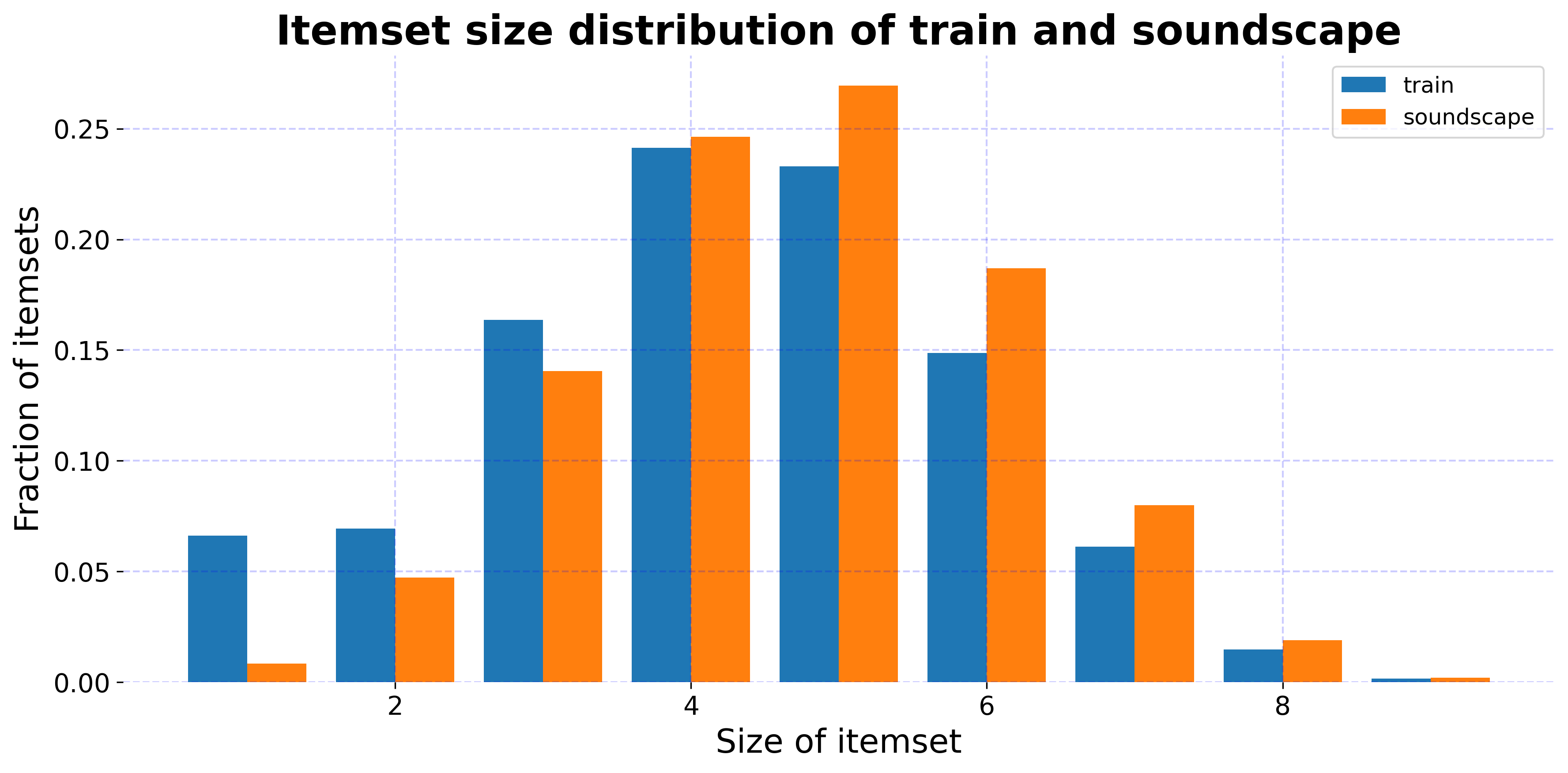}
    \caption{
        The plot shows the distribution of itemset sizes in the training and soundscape datasets.
        The distribution represents how likely species are to co-occur in each recording.
    }
    \label{fig:itemset-size}
\end{figure}

We use a frequent-pattern mining algorithm, FPGrowth \cite{han2000mining}, to identify co-occurrence patterns in the soundscape dataset.
In Table \ref{table:itemset_data}, we observe that co-occurrences of species can appear more often than individual species alone.
The frequent itemsets give us a rough estimate of how many birds we can expect to see in a single recording.
In Figure \ref{fig:itemset-size}, we plot the distribution of normalized itemset sizes in the training and soundscape datasets.
We observe an approximately normal distribution of sizes centered around four to six species per recording.
The training set is skewed toward smaller itemsets, likely due to biases in the data collection process where individuals are more likely to record and upload isolated calls.

\section{Methodology}

The experiments are run over several modular stages.
We implement an end-to-end workflow that applies domain-specific fine-tuning to state-of-the-art models for birdcall classification.
We quantify differences between choices of dataset, architecture, and training losses.
The second part of the experiment focuses on transfer learning using the model as a surrogate, where the model's predictions are used as labels for transfer learning on audio classification models.
In particular, we study a widely distributed birdcall-specific convolutional neural network and a self-supervised neural audio codec for encoding and decoding.

\subsection{Transfer Learning}

The Google Bird Vocalization Classification model is the main surrogate transfer learning experiment, focused on version 4 of \texttt{google/bird-vocalization-classifier} on Kaggle, corresponding to version 1.3 on the TensorFlow hub.
We directly compute a prediction for the competition by selecting the competition subset, filling in the missing species with negative infinity, and computing the sigmoid of the logits.
We perform fine-tuning of the model by training a new classification head on the training dataset using the thresholded predictions of the model as pseudo-labels for the multi-label classification task.
We take advantage of the species label of the folder according to Section \ref{ssec:pseudo_multi_label} as one form of augmentation.
We also fine-tune the model on the unlabeled soundscape data.

\begin{figure}[ht]
  \centering
  \includegraphics[width=1\linewidth]{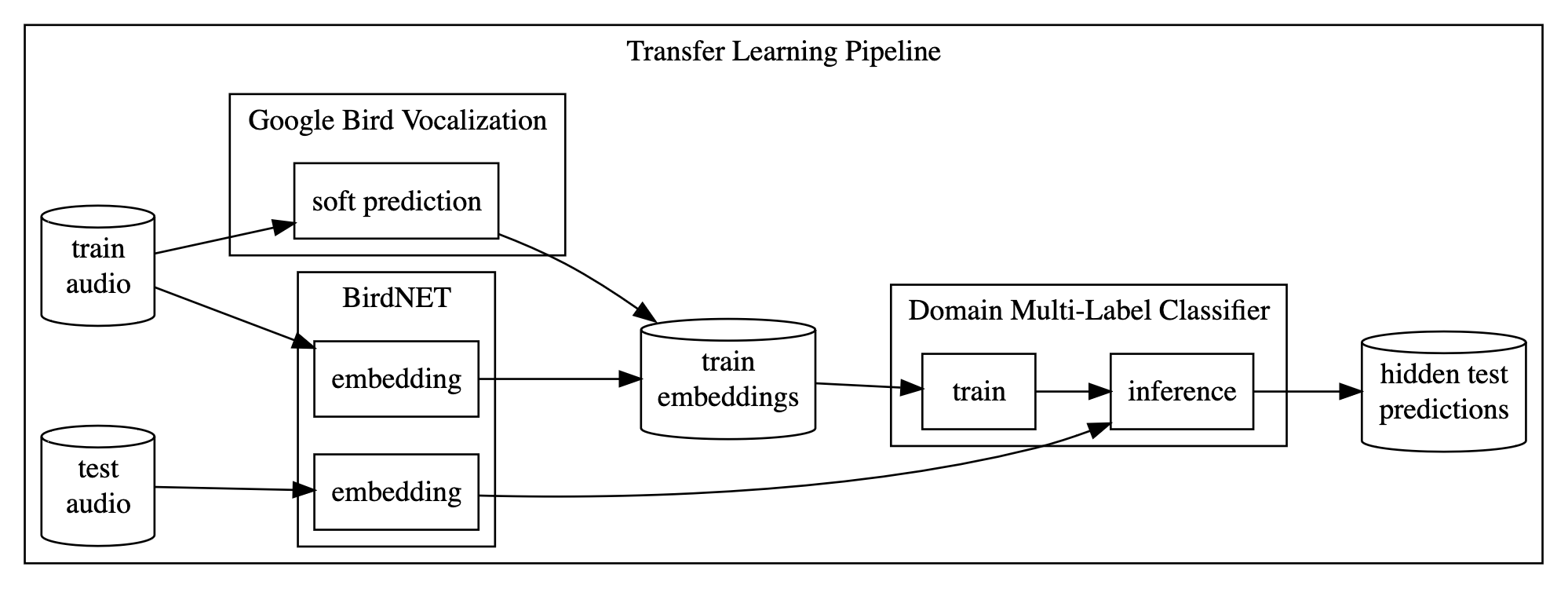}
  \caption{
    Diagram of the transfer learning pipeline used for experiments with BirdNET and Encodec with the Google Bird Vocalization model as a surrogate.
    The soft predictions from the Bird Vocalization model are used as pseudo-labels
    to train a multi-label classifier on BirdNET's embedding space.
    BirdNET is also replaced with EnCodec for comparison.
  }
  \label{fig:pipeline}
\end{figure}

\break
We experiment with different losses to optimize the multi-label classifier, including binary cross-entropy (BCE), asymmetric loss (ASL), and sigmoidF1 which are explained in Section \ref{ssec:training_losses}.
We experimented with a hidden layer to increase the capacity of the models.
The model is trained for 20 epochs with a batch size of 1,000 and a learning rate calculated by PyTorch Lightning.
The training dataset is split with an 80-20 train-validation split.
The model is trained on a single NVIDIA L4 on a Google Cloud Platform (GCP) g2-standard-8 instance with 8 vCPU, 16GB of memory, and 375GB of local NVME storage for dataset caching and model checkpoints.

We use BirdNET V2.4 through \texttt{joeweiss/birdnetlib} and EnCodec through \texttt{facebookresearch/encodec} v0.1.1 as comparisons for knowledge transfer.
Though both the Bird Vocalization model and BirdNET provide predictions for classification, the former provides a more extensive set of species that overlaps with this year's competition.
We ignore the outputs of the BirdNET model for our experiments and focus on learning the distribution of the Bird Vocalization model's outputs.

\subsection{Data Preprocessing}

We pre-compute the embeddings and predictions of the Bird Vocalization model on the training and unlabeled soundscape datasets into a binary, columnar format that is easily accessible from network storage.
The embeddings are in a $\mathcal{R}^{1280}$ space, while predictions are limited to the competition's species set.
If the species is not present, its prediction is set to zero by assigning negative infinity to the logit output.
To save computation, we also pre-compute and join the embeddings from BirdNET and EnCodec with the predictions of the Bird Vocalization model.

For BirdNET, we must align the model's input size of 48kHz of 3 seconds to the 32kHz of 5 seconds that both the Bird Vocalization model and BirdCLEF competition expect.
We take the mean of the embeddings of the 5-second audio clip with a 1-second stride for the 0th and 2nd seconds.
This provides coverage of the entire audio clip while limiting the computational burden of encoding.
We take 5-second embedding tokens at 24kHz, and limit the bandwidth of Encodec to 1.5kbps for an embedding space of $\mathcal{R}^{5\times150}$.
Increasing the bandwidth to 3kpbs leads to an embedding space of $\mathcal{R}^{5\times300}$.
We qualitatively inspect the embeddings through a cluster analysis in Figure \ref{fig:pacmap-cluster}, noting the relative difficulty of separating common classes within the dataset.

\subsection{Pseudo Multi-Label Construction}
\label{ssec:pseudo_multi_label}

The training dataset lacks traditional labels for supervised learning, as the 5-second intervals in each recording are not labeled with the species present.
We use pseudo-labels derived from the thresholded predictions of a surrogate model, which are not human-verified ground truth.
Additionally, we use the folder species as an extra label for further training the model.
The thresholded predictions are defined as a function of the model's output $\hat{y}$ and a threshold $p_{\text{threshold}}$, with the sigmoid function denoted by $\sigma$.

\begin{equation}
  \hat{y} = \sigma(\text{g}(x)) > p_{\text{threshold}}
\end{equation}

We define an indicator variable $\mathbbm{1}_{\text{call}}$ that determines whether the model output detects a birdcall, which occurs when any species prediction is positive.

\begin{equation}
  \mathbbm{1}_{\text{call}}(x) = \sum_{i=0}^{\left|x
  \right|}x_i > 0
\end{equation}

We also generate a one-hot encoding of the folder that the current audio belongs to $\mathbbm{1}_{\text{species}}$, where $S$ is the set of species in the folder.

\begin{equation}
  \mathbbm{1}_{\text{species}}(x) = \begin{cases}
    1 & \text { when } x \in S \\ 
    0 & \text { when } x \notin S
  \end{cases}
\end{equation}

Finally, we can define our modified label as the intersection of the model's output and the species of the folder.
This can be implemented as a vectorized operation in PyTorch.

\begin{equation}
  \hat{y}_{\text{species}} = \hat{y}\ \lor\ (\mathbbm{1}_{\text{call}}(\hat{y}) \land \mathbbm{1}_{\text{species}}(s)) 
\end{equation}

We use a threshold of $p_{\text{threshold}} = 0.5$ when defining all labels.
Experiments on the unlabeled soundscapes do not have the additional information provided in the training dataset, and thus we are limited to the pseudo-labels $\hat{y}$.

\subsection{Training Losses} 
\label{ssec:training_losses}

We experiment with different losses to optimize the multi-label classifier.
The competition evaluation uses a modified ROC-AUC that skips classes with no true-positive labels.
We utilize \texttt{MultiLabelAUROC} from the \texttt{torchmetrics} library as the primary learning metric.
We also consider the macro-F1 score as a secondary metric, which was utilized in the 2022 edition of the BirdCLEF competition \cite{kahl2022overview}.
This metric allows us to inspect other aspects of the loss functions we consider in our experiments.

\subsubsection{Binary Cross-Entropy}

Binary cross-entropy is a loss function used for binary classification.
It is suitable for multi-label classification as it treats each label as an independent binary classification problem.
We use this loss as a baseline due to its simple interpretation and absence of hyperparameters.

\begin{equation}
L = -\sum_{c=1}^My_{o,c}\log(p_{o,c})
\end{equation}

\subsubsection{Asymmetric Loss (ASL)}

The asymmetric loss \cite{ridnik2021asymmetric} penalizes false positives and false negatives differently.
This construction dynamically down-weights easy negative samples, hard thresholds them, and ignores misclassified samples.
This loss is well-suited for our problem domain since we have fuzzy labels from another model initially intended for single-label classification.

\begin{equation}
  ASL=\left\{
    \begin{array}{l}
      L_{+}=(1-p)^{\gamma_{+}} \log (p) \\
      L_{-}=\left(p_m\right)^{\gamma_{-}} \log \left(1-p_m\right)
    \end{array}\right.
\end{equation}

The loss is defined in terms of the probability of the network output $p$ and hyper-parameters $\gamma_{+}$ and $\gamma_{-}$.
Setting $\gamma_{+} > \gamma_{-}$ emphasizes positive examples while setting both terms to 0 yields binary cross entropy. 
We sweep over parameters $\gamma_{+} \in \{0, 1\}$ and $\gamma_{-} \in \{0, 2, 4\}$, while the default values are $\gamma_{+}=1$ and $\gamma_{-}=4$,

\subsubsection{sigmoidF1}

The sigmoidF1 loss \cite{benedict2021sigmoidf1} optimizes the F1 score directly by creating a differentiable approximation of the F1 score.
Though the competition does not score with F1, it provides a useful point of comparison with other losses.
We first define the true positive, false positive, false negative, and true negative terms as a function of the sigmoid function.

\begin{equation}
  \begin{aligned}
  & \widetilde{t p}=\sum \mathbf{S}(\hat{\mathbf{y}}) \odot \mathbf{y} \quad
  \tilde{f p}=\sum \mathbf{S}(\hat{\mathbf{y}}) \odot(\mathbbm{1}-\mathbf{y})
  \\
  & \tilde{f n} =\sum(\mathbbm{1}-\mathbf{S}(\hat{\mathbf{y}})) \odot \mathbf{y} \quad
  \tilde{t n}=\sum(\mathbbm{1}-\mathbf{S}(\hat{\mathbf{y}})) \odot(\mathbbm{1}-\mathbf{y})
  \end{aligned}
\end{equation}

where $\mathbf{S}(\hat{\mathbf{y}})$ is the sigmoid function applied to the model's output $\hat{\mathbf{y}}$.

\begin{equation}
  S(u ; \beta, \eta)=\frac{1}{1+\exp (-\beta(u+\eta))}
\end{equation}

Then we define the F1 score as a function of the true positive, false positive, and false negative terms.

\begin{equation}
  \mathcal{L}_{\widetilde{F 1}}=1-\widetilde{F 1}, \quad \text { where } \quad \widetilde{F 1}=\frac{2 \widetilde{t p}}{2 \widetilde{t p}+\widetilde{f n}+\widetilde{f p}}
\end{equation}

We are given two hyper-parameter $S=-\beta$ and $E=\eta$.
We sweep over parameters $S \in \{-1, -15, -30\}$ and $E \in \{0, 1\}$ as suggested in the author's experiments.



\section{Results}

We obtain results for various models on the leaderboard via code submission on Kaggle.
We report the best validation F1 and AUROC scores, together with private and public leaderboard scores.
All submissions were made past the competition deadline with the exception of the starter Keras notebook.
We submit a model that predicts 0 for every species on the leaderboard, leading to a private and public score of 0.5.
We submit the predictions from the Bird Vocalization model and obtain a private and public score of 0.516625 and 0.556097 respectively.

\subsection{Loss Comparisons}

In Table \ref{tab:vocalization-summary}, we train a linear classifier head against combinations of BCE and ASL with the addition of the species label logic.
We report our validation F1 and AUROC scores alongside the private and public scores.
We note that AUROC quickly saturates against the validation set used in the training dataset.
The validation F1 score however correlates more strongly with the leaderboard scores.
Using the species labels typically increases the score by ~0.05 e.g. ASL with default parameters goes from 0.529 to 0.576 in the public leaderboard.

\begin{table}[h!]
\caption{
Overview of linear classifier heads on Bird Vocalization embeddings using the train dataset.
}
\label{tab:vocalization-summary}
\begin{tabular}{|l|l|p{0.6in}|p{0.6in}|p{0.5in}|p{0.5in}|}
\hline
\textbf{Loss} & \textbf{Description} & \textbf{Val F1} & \textbf{Val AUROC} & \textbf{Private Score} & \textbf{Public Score} \\ \hline
BCE & labels $\hat{y}$ & 0.5420 & 0.999 & 0.508008 & 0.538838 \\ \hline
BCE & labels $\hat{y}_{species}$ & 0.6095 & 0.997 & 0.546232 & 0.583395 \\ \hline
ASL & labels $\hat{y}$  & 0.6035 & 0.999 & 0.523207 & 0.529498 \\ \hline
ASL & labels $\hat{y}_{species}$ & 0.6603 & 0.998 & 0.556189 & 0.576463 \\ \hline
\end{tabular}
\end{table}

We experimented with adding a hidden layer behind the classification head to encourage the model to learn more complex patterns.
Using ASL as the loss function, we varied the hyperparameters listed in Table \ref{tab:asl-summary}.
We confirmed the efficacy of the species logic but noted that the scores were marginally lower than those of the linear models.
Additionally, we found that the default parameters of ASL are effective in most tasks, with minimal tuning needed for good performance on domain-specific tasks. 

\begin{table}[h!]
\caption{
  An overview hyper-parameter tuning ASL on a $\mathcal{R}^{256}$ hidden-layer model.
  Parameters of $\gamma_{-}=2$ and $\gamma_{+}=1$ tend to work the best.
}
\label{tab:asl-summary}
\begin{tabular}{|l|l|p{0.6in}|p{0.6in}|p{0.5in}|p{0.5in}|}
\hline
\textbf{Description} & \textbf{Hyperparameters} & \textbf{Val F1} & \textbf{Val AUROC} & \textbf{Private Score} & \textbf{Public Score} \\ \hline
labels $\hat{y}$            & \makecell{$\gamma_{-} = -2, \gamma_{+} = 1$} & 0.5727 & 0.996 & 0.557257 & 0.538430 \\ \hline
labels $\hat{y}$            & \makecell{$\gamma_{-} = -4, \gamma_{+} = 0$} & 0.5731 & 0.996 & 0.521133 & 0.520745 \\ \hline
labels $\hat{y}$            & \makecell{$\gamma_{-} = -4, \gamma_{+} = 1$} & 0.5758 & 0.996 & 0.545155 & 0.524194 \\ \hline
labels $\hat{y}_{species}$  & \makecell{$\gamma_{-} = -2, \gamma_{+} = 1$} & 0.6563 & 0.997 & 0.585699 & 0.556193 \\ \hline
labels $\hat{y}_{species}$  & \makecell{$\gamma_{-} = -4, \gamma_{+} = 0$} & 0.6414 & 0.997 & 0.558255 & 0.534317 \\ \hline
labels $\hat{y}_{species}$  & \makecell{$\gamma_{-} = -4, \gamma_{+} = 1$} & 0.6495 & 0.997 & 0.542350 & 0.558353 \\ \hline
\end{tabular}
\label{tab:twolayer-summary}
\end{table}

\subsection{Embedding Model Comparisons}

We summarize the performance of each loss function across the CNN-based models in Table \ref{tab:loss-summary}.
Due to CPU-time limitations on notebook runtime, we do not include an EnCodec-based model.
Our best model on the public leaderboard uses BirdNET embeddings and the BCE loss.
BirdNET embeddings consistently perform better with linear models, despite the origin of the labels being the Bird Vocalization model.
Access to the species label from the parent folder consistently improves scores.
While BCE performs well, this behavior is not indicated by our validation and private test metrics alone.

\begin{table}[h!]
  \caption{
  A comparative overview of Bird Vocalization and BirdNET linear models with different losses and labeling logic on the training dataset.}
  \label{tab:loss-summary}
  \begin{tabular}{|l|ll|ll|}
    \hline
    \multirow{2}{*}{\textbf{Model}} & \multicolumn{2}{l|}{\textbf{No Species}}                & \multicolumn{2}{l|}{\textbf{With Species}}              \\ \cline{2-5} 
                                    & \multicolumn{1}{l|}{\textbf{Private}} & \textbf{Public} & \multicolumn{1}{l|}{\textbf{Private}} & \textbf{Public} \\ \hline
    Bird Vocalization (BCE)  & \multicolumn{1}{l|}{0.508008} & 0.538838 & \multicolumn{1}{l|}{0.546232} & 0.583395 \\ \hline
    Bird Vocalization (ASL)  & \multicolumn{1}{l|}{0.523207} & 0.529498 & \multicolumn{1}{l|}{0.556189} & 0.576463 \\ \hline
    Bird Vocalization (sigmoidF1)        & \multicolumn{1}{l|}{0.529473}         & 0.553349        & \multicolumn{1}{l|}{0.566378}         & 0.596767        \\ \hline
    BirdNET (BCE)       & \multicolumn{1}{l|}{0.526441} & 0.577005 & \multicolumn{1}{l|}{0.562368} & 0.630415 \\ \hline
    BirdNET (ASL)       & \multicolumn{1}{l|}{0.505578} & 0.538719 & \multicolumn{1}{l|}{0.550123} & 0.599697 \\ \hline
    BirdNET (sigmoidF1) & \multicolumn{1}{l|}{0.543268} & 0.576497 & \multicolumn{1}{l|}{0.559693} & 0.590848 \\ \hline
  \end{tabular}
\end{table}

\subsection{Dataset Comparisons}

In Table \ref{tab:soundscape-summary}, we compare the performance of linear models trained on the soundscape dataset using ASL as the main loss.
We observe two main results: (1) BirdNET embeddings outperform the bird vocalization model by 0.03 on the public leaderboard and (2), models trained on the soundscape dataset are less effective than those trained on the distribution of the training dataset.
This may be attributed to ASL's dynamic downscaling of easily classified negative labels, making the contribution of training labels more significant than the similarity to the test distribution.

\begin{table}[h!]
\caption{
    Scores for transfer learning on soundscapes using ASL.
}
\label{tab:soundscape-summary}
\begin{tabular}{|l|p{0.6in}|p{0.6in}|p{0.5in}|p{0.5in}|}
\hline
\textbf{Description} & \textbf{Val F1} & \textbf{Val AUROC} & \textbf{Private Score} & \textbf{Public Score} \\ \hline
Bird Vocalization, Linear, ASL ($\gamma_{-}=2, \gamma_{+}=1$)  & 0.1423 & 0.999 & 0.448845 & 0.499735 \\ \hline
BirdNET, Linear, ASL ($\gamma_{-}=2, \gamma_{+}=1$) & 0.1158 & 0.997 & 0.47927  & 0.532035 \\ \hline
\end{tabular}
\label{tab:twolayer-summary}
\end{table}

\begin{table}[h!]
\caption{
    Profiling information for inference of various models, using the Python profiler via Lightning.
    Timing information is collected from the \texttt{predict\_next} step on the first 20 soundscapes sorted by identifier.
    The rate of inference time per soundscape allows extrapolation to the hidden test set of 1100 4-minute recordings.
}
\label{tab:profiling}
\begin{tabular}{|l|r|r|r|}
\hline
Name & Profile (sec) & Rate (sec/4m) & Test Estimate (hours) \\ \hline
torchaudio                                      & 1.1   & 0.05  & 0.02 \\ \hline
vocalization passthrough noncompiled & 188.6            & 9.43                  & 2.88                                    \\ \hline
vocalization passthrough compiled& 24.0  & 1.20  & 0.37 \\ \hline
vocalization linear compiled     & 64.6  & 3.23  & 0.99 \\ \hline
birdnet passthrough compiled     & 56.9  & 2.85  & 0.87 \\ \hline
encodec passthrough noncompiled  & 156.4 & 7.82  & 2.39 \\ \hline
encodec passthrough compiled     & 213.7 & 10.69 & 3.27 \\ \hline
\end{tabular}
\end{table}

\subsection{Inference Runtime}

We profile each model to estimate the time required to process all test soundscapes, as shown in Table \ref{tab:profiling}.
The Python profiler measures the time spent in each function and the number of function calls.
Reading all audio into chunked arrays from disk into memory, our baseline takes approximately one minute.

The Bird Vocalization model did not complete within the Kaggle's time constraints, taking nearly three hours according to our estimates.
We compile the model using TensorFlow Lite at runtime, optimizing operations for the hardware while allowing fallback to non-lite operations.
This compilation process results in an order-of-magnitude performance increase, leaving a substantial margin for additional computation.
The linear classification head adds only an extra half-hour of computation.
The BirdNET model also runs well within time constraints as it is compiled with TensorFlow Lite.

EnCodec exceeds the time budget, taking 2.4 hours for the base model.
Experimenting with OpenVINO \cite{openvino2019} and applying data-independent quantization and compression did not improve inference speed.
\section{Discussion}

\subsection{Transfer Learning Experimentation}

Our transfer learning experiments using the Bird Vocalization classifier exhibit different behaviors between the private and public leaderboards.
While fine-tuned models outperform the base model when trained on the subset of species provided for the competition, we hypothesize a shift in the species distribution between the private and public test sets.
The Bird Vocalization model is trained on a more balanced dataset drawn from a larger set of species, whereas our transfer learning techniques rely on pseudo-labeling from the donor model, which may not be well-calibrated for this task.
We did not account for the skew in the training data, apparent from the distribution of audio of each species.

We address label skew through different loss function choices.
We use a secondary metric during training to provide another axis to compare models.
When fine-tuning the Bird Vocalization classifier to learn the outputs from the original classifier head, the AUROC loss converges close to unity across various architectures.
However, different losses exhibit varying learning behaviors against the F1-score, with some designed to be better surrogates than binary cross-entropy loss.
During transfer learning, these losses provide a smooth, monotonic increase to the validation F1-score, indicating that Bird Vocalization embeddings offer a "good" representation of domain-specific data for the multi-label problem.
We observe different behaviors in other embedding spaces, supported by our clustering charts.

\begin{figure}[h!]
    \centering
    \includegraphics[width=1\linewidth]{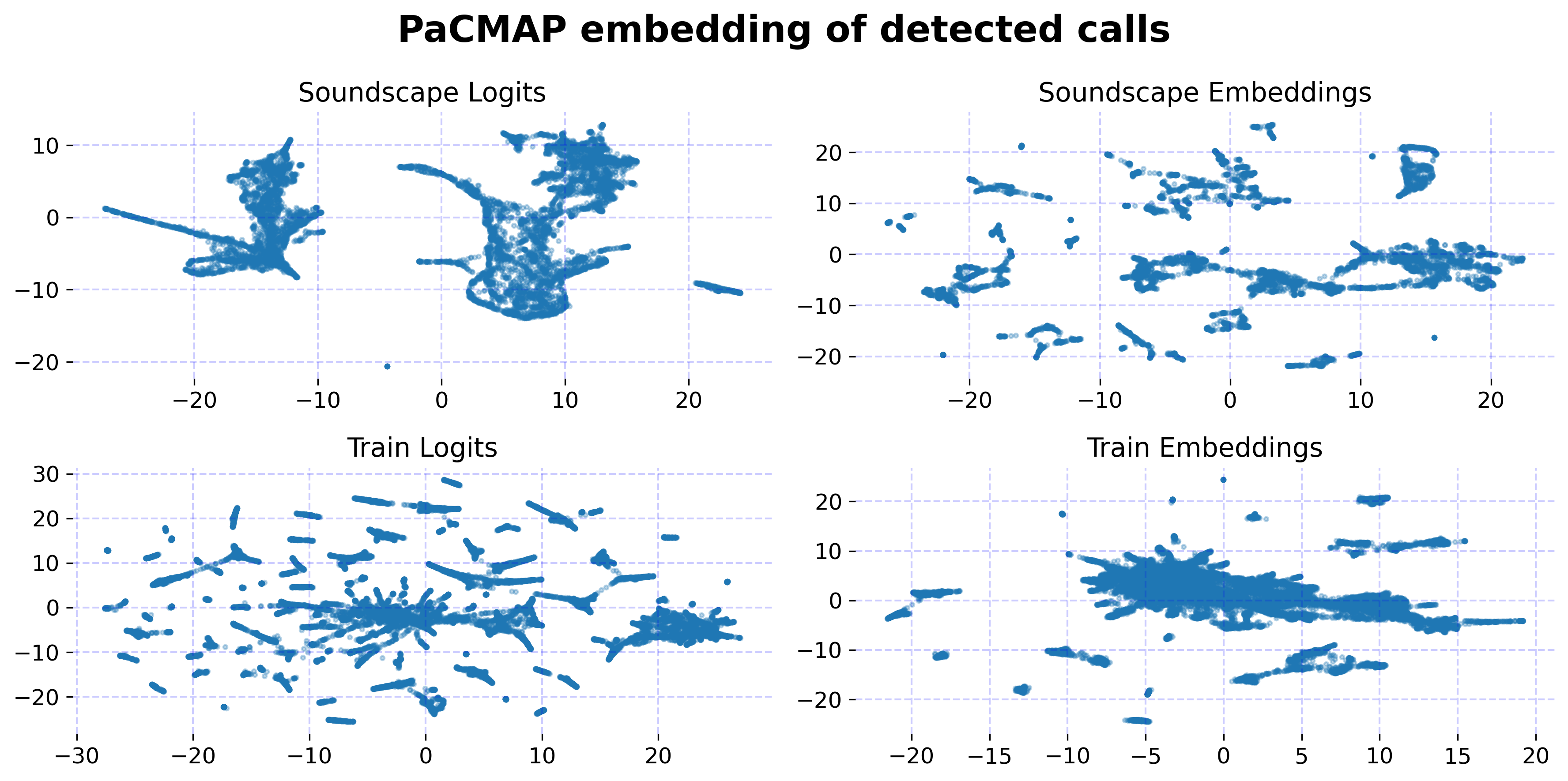}
    \caption{
        A clustering analysis of the embeddings and logits extracted from the Bird Vocalization Classifier, on the training and soundscape data.
        We obtain a single vector for each track by taking the max value of the logits and the mean value of the embeddings.
        The resulting vectors are clustered using PaCMAP and demonstrate distinctive topology resulting from distinct distributional semantics.
    }
    \label{fig:detected-calls}
\end{figure}

To address skew in the training dataset, the organizers provide unlabeled soundscapes representative of the hidden-test dataset.
We discuss the distributional shift between species and frequent itemsets in Section \ref{sec:eda}.
Figure \ref{fig:detected-calls} shows the active intervals of calls, revealing differences in data geometry.
The train datasets at the bottom have tightly clustered logits, likely representing peaks in species probability distributions.
The embeddings form a large central cluster with several outliers, probably representing distinctive calls.
Conversely, the soundscape logit space forms two major clusters, reflecting the smaller set of species present.
Thus, soundscape embeddings should closely reflect clusters of birdcalls.
It would be interesting to explore how well we can discriminate between recording sites, as location likely correlates with species distribution and co-occurrence patterns.

We expect soundscapes to better represent the species distribution in the hidden test set.
However, our results show that the models trained in the soundscapes perform worse than those trained on the original dataset.
Although the addition of soundscapes adds an interesting dimension to the competition, it requires more than cursory experimentation to incorporate into modeling effectively.

\subsection{Self-Supervised Neural Codecs}

We find that EnCodec does not transfer well with similar experiments involving the linear and two-layer classifiers, achieving validation F1-scores below 0.1.  
Adding an LSTM layer to handle the sequential nature of EnCodec embeddings did not improve the scores. 
A much deeper model, similar to the EnCodec decoder \cite{defossez2022encodec}, is likely needed to learn from the quantized embeddings, but this is not feasible within the competition's inference time constraints.

Additionally, EnCodec is computationally expensive and difficult to adapt to the constrained submission environment.
The Python profiler identified model inference as the bottleneck, with most time spent on EnCodec inference.
OpenVINO post-training optimizations for quantizing and compressing weights do not significantly improve inference throughput, likely due to existing optimizations in the upstream library.
A $1.5 \times$ speedup is needed to use EnCodec in our pipeline, indicating that further optimizations are required to leverage neural codecs based on large datasets trained with attention and self-supervision.

\section{Future Work}

\begin{figure}[h!]
    \centering
    \includegraphics[width=0.9\linewidth]{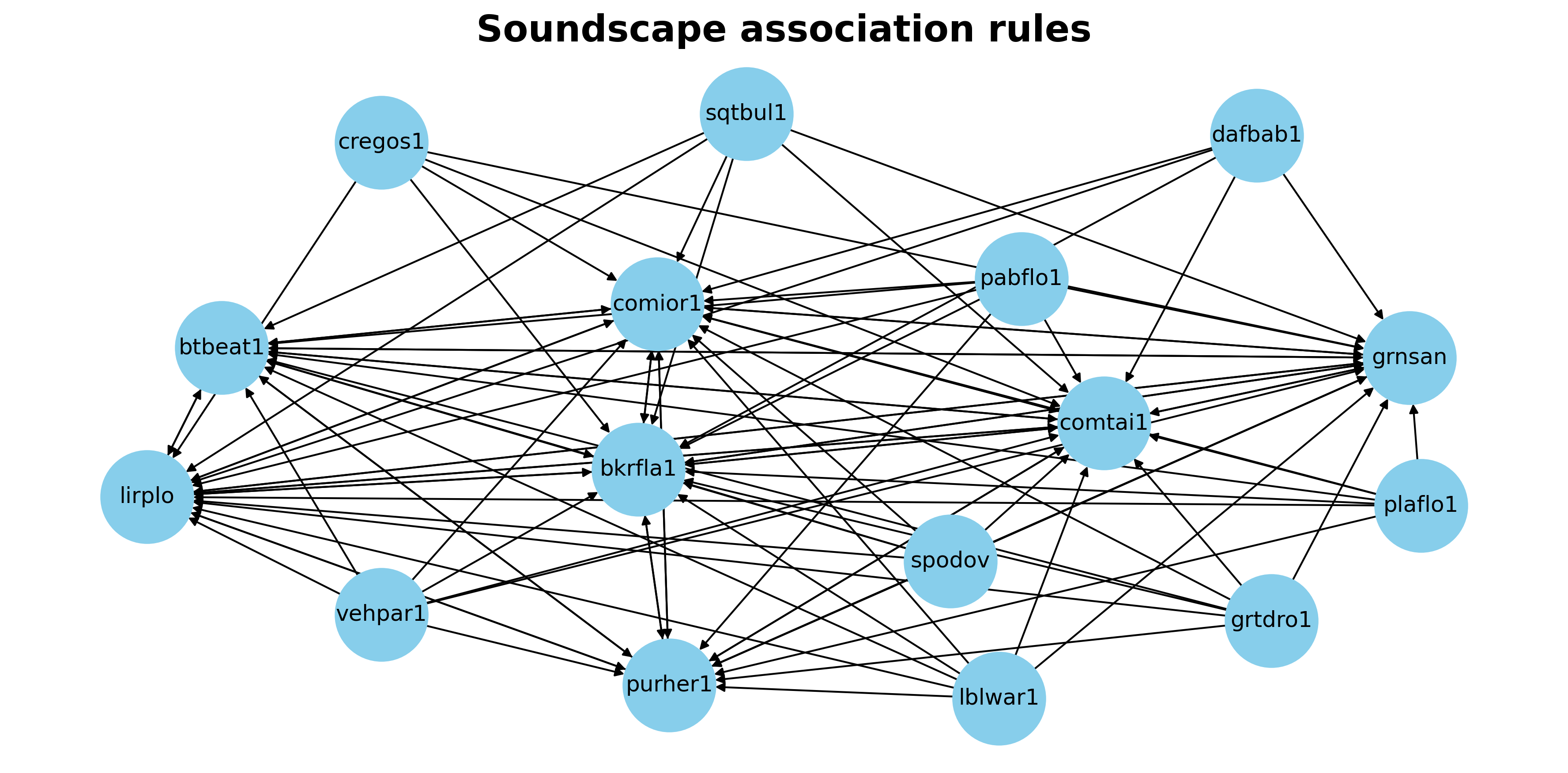}
    \caption{
        FPGrowth generates association rules from the frequent itemsets using the default minimum confidence threshold of 0.8.
        Consequent itemsets are obtained with an individual species as the antecedent, and a directed graph is created by drawing edges between items in the antecedent and consequent.
    }
    \label{fig:soundscape-rules}
\end{figure}

Exploiting co-occurrence species information as a prior to the learning process could be beneficial.
We have demonstrated frequent pattern mining to obtain co-occurrence distributions and quantify differences from the training dataset.
Confident relationships extracted from the data can be visualized, as shown in Figure \ref{fig:soundscape-rules}, and used to reshape the probability distribution of an existing classifier to better represent the posterior of the unlabeled soundscape.

We aim to explore alternative parameterizations of sequential models that are computationally viable for future competitions.
The competition's trade-offs favor compact domain-specific models over large neural networks, focusing on linearithmic algorithms like the Fast Fourier transform for input representation.
Finding a pre-trained neural audio codec with fewer parameters that fit within our computational budget and pass human perceptual tests could be viable.
Alternatively, training models from scratch using different architectures via distillation methods, compatible with the encoder-decoder architecture used in EnCodec, could be explored.
State-space models like Mamba \cite{gu2023mamba} provide an appealing alternative to attention-based methods, potentially staying within our computational budget.

\section{Conclusion}

Our study demonstrates the effectiveness of transfer learning in birdcall classification using embeddings from pre-trained models like Google's Bird Vocalization Classification Model and BirdNET. 
These embeddings capture meaningful structures that are beneficial for multi-label classification, although they do not outperform many top models in the competition.
Our best-performing model, which uses BirdNET embeddings and Bird Vocalization pseudo labels to train a linear classifier, achieved a 0.63 score on the post-competition public leaderboard.
Future work will focus on optimizing computational efficiency and exploring alternative model architectures to better handle the sequential nature of audio data. 
We also plan to incorporate species co-occurrence patterns to further enhance classification accuracy. 
Our code is available at \href{https://github.com/dsgt-kaggle-clef/birdclef-2024}{github.com/dsgt-kaggle-clef/birdclef-2024}.

\begin{acknowledgments}
Thank you to the Data Science at Georgia Tech (DS@GT) club for providing hardware for experiments, and to the organizers of BirdCLEF and LifeCLEF for hosting the competition.

\end{acknowledgments}

\bibliography{report.bib}

\end{document}